\documentclass[trackchanges]{aastex701}
\usepackage{todo}
\usepackage{comment}

\begin{document}
\title{Exploring Case B I: The Footprints of Hidden Radiation Fields in H~I Balmer Decrements}

\author[orcid=0000-0000-0000-0001,sname='Dehghanian']{Maryam Dehghanian}
\affiliation{Physics \& Astronomy, University of Kentucky, Lexington, Kentucky, USA}
\email[show]{m.dehghanian@uky.edu}  

\author[orcid=0000-0002-2915-3612,sname=Guzm\'an]{F. Guzm\'an}
\affiliation{Physics \& Astronomy, University of North Georgia, Dahlonega, Georgia, USA }
\email[hide]{francisco.guzmanfulgencio@ung.edu} 

\author[orcid=0000-0001-7490-0739,sname=van Hoof]{P. A. M. van Hoof}
\affiliation{Royal Observatory of Belgium, Ringlaan 3, 1180 Brussels, Belgium}
\email[hide]{p.vanhoof@oma.be} 

\author[orcid=0000-0002-4634-5966,sname=Gunasekera]{C. Gunasekera}
\affiliation{Space Telescope Science Institute, Baltimore, MD, USA}
\email[hide]{cgunasekera@stsci.edu}

\author[orcid=0000-0003-4503-6333,sname='Ferland']{Gary Ferland} 
\affiliation{Physics \& Astronomy, University of Kentucky, Lexington, Kentucky, USA}
\email[hide]{m.dehghanian@uky.edu} 


\begin{abstract}
Theory predicts that H~I emission from photoionized clouds should
lie close to the Case~B limit established by Menzel and Baker, and observations
of nearby nebulae have largely confirmed this expectation. However, recent
observations of more distant high-redshift systems have revealed a small fraction with striking
departures from Case~B. What additional physics might be at work in these
objects?

This paper is part of a series exploring the physical processes that can
produce such departures. In Paper~I (D26), we
show that Case~B remains remarkably robust over a broad range of physical
conditions, consistent with observations of nearby nebulae. Here, we explore
one additional ingredient: a soft, non-ionizing radiation field. Such a continuum can strongly perturb the Balmer spectrum. The largest effects
occur for radiation fields resembling $\log(T_{\rm eff}/{\rm K})\sim3.2$--$4.2$
blackbodies, where photoexcitation from the metastable $2s$ level alters the
hydrogen level populations and subsequent radiative cascade. The resulting
H$\alpha$/H$\beta$ and H$\gamma$/H$\beta$ ratios follow a different trajectory
from dust reddening, allowing the two effects to be distinguished using
multiple Balmer lines. In unresolved systems, a cool source located near the line-emitting gas may provide an additional radiation field distinct from the hot stars responsible for ionization. Balmer decrements may therefore
carry information about otherwise hidden components of the radiation field
surrounding the emitting gas.
\end{abstract}



\section{Introduction}
Hydrogen recombination lines are among the most widely used diagnostics in astrophysics. A foundational series of papers by Menzel and Baker \citep{1937ApJ....85..330M,1937ApJ....86...70M,1938ApJ....88...52B,1938ApJ....88..422B} established the physical limits that govern these lines in photoionized gas. They recognized that 
the emergent hydrogen spectrum depends not only on the atomic 
recombination itself, but also on the fate of resonant photons and
the gas's interaction with the surrounding radiation field. 
They introduced a set of limiting cases---now known as Cases~A, B, and C
\citep[][Herafter AGN3]{1937ApJ....85..330M,1937ApJ....86...70M,1938ApJ....88...52B,1938ApJ....88..422B,2021ApJ...912...26C, 2006agna.book.....O} ---that describe different regimes of radiative transfer and 
excitation. 


Lines produced in Cases A and C are far fainter than Case B since
the Lyman continuum is optically thin and little ionizing radiation is reprocessed.
For this reason, Case~B recombination has become the standard 
theoretical framework for interpreting hydrogen emission-line ratios in nebular astrophysics. In Dehghanian et al. (2026, in preparation), hereafter D26, the first paper in this series, we explore the Case~B limit and its underlying physics in greater detail. D26 found that Case~B applies across a wide range of H II regions and planetary nebulae.

Recently, several observations have found non-Case~B hydrogen emission in a variety of astrophysical systems
and proposed a variety of explanations, ranging from changes in the ionizing
continuum to Lyman-line pumping, Balmer self-absorption, and other
radiative-transfer effects \citep[see, e.g.,][]{2023ApJ...950..106W,
2024arXiv240409015S,2024ApJ...974..180Y,2025MNRAS.540..190M,
2025arXiv251006362N,2025arXiv250321896S}.

Motivated by these observations, we have begun a systematic
investigation of the physical processes capable of driving hydrogen emission
away from the canonical Case~B limit.

\begin{figure*}[ht!]
\plotone{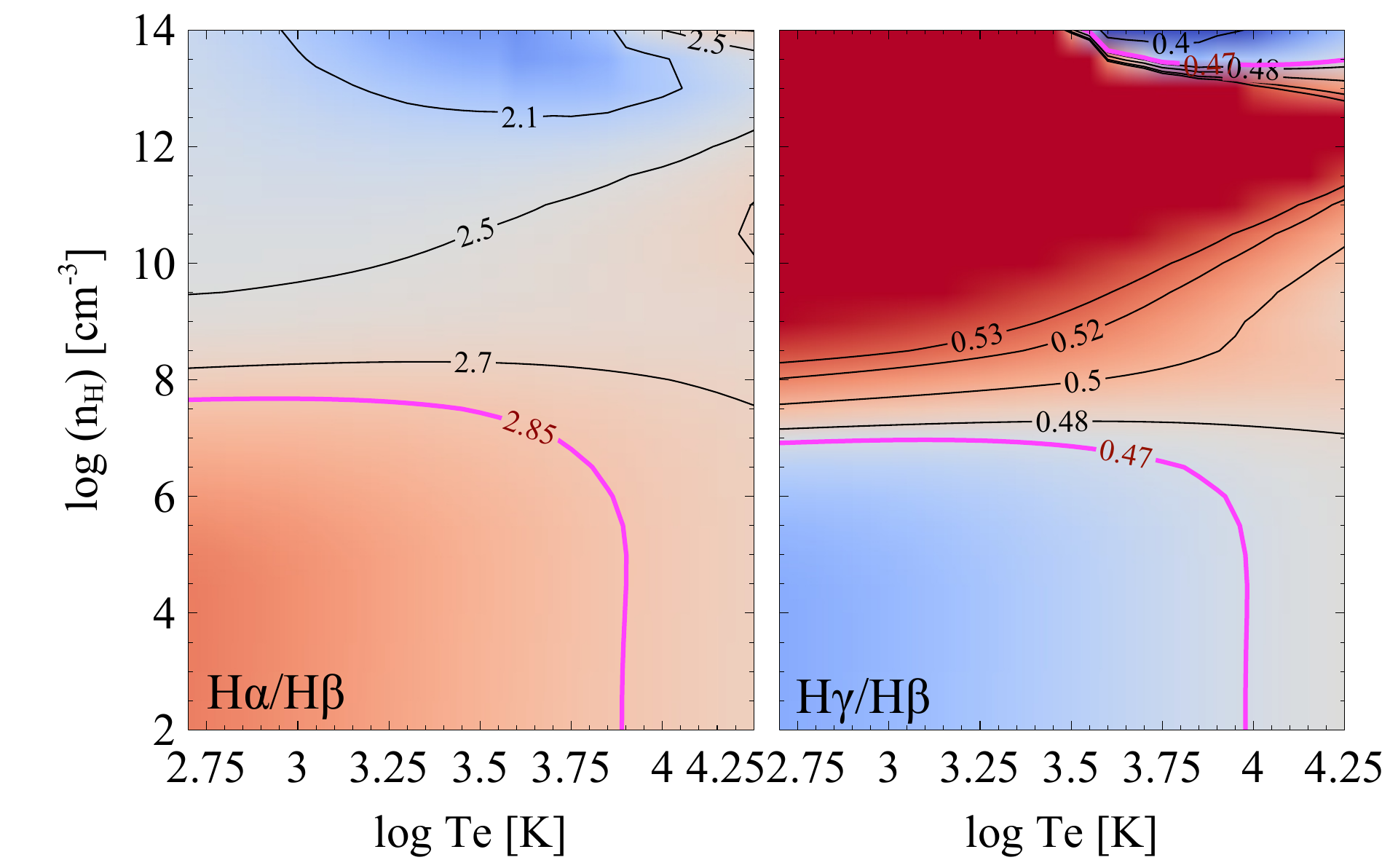}
\caption{Case~B H$\alpha$/H$\beta$ (left) and H$\gamma$/H$\beta$ (right)
ratios as functions of electron temperature and hydrogen density. The magenta
contours show the low-density Case~B values for a temperature
of 10$^4$ K. The setup is described near figure 6 of
\citet{2017RMxAA..53..385F}.}
\label{fig:orig}
\end{figure*}

The most extensive studies of H~I emission have focused on the
three brightest Balmer lines, H$\alpha,\ \beta,\ $and$\ \gamma$.
Figure~\ref{fig:orig} shows Case B calculations using
the development version of Cloudy, last described by
\citet{2025RMxAA..61..120G}. For these calculations, we extended the H~\textsc{i} model atom to include
15 resolved and 100 collapsed levels, allowing the radiative cascade and the
response of the excited-state populations to the additional radiation field
to be followed in greater detail.  The setup is very
similar to the Case B simulations in fig.~6 of \citet{2017RMxAA..53..385F}.
Normally, Cloudy determines
the kinetic temperature, density, and optical depths of all
lines self-consistently. In this case we set the temperature,
vary the hydrogen density, and set the optical depths in all the Lyman
lines to a large value. This provides a convenient way
to test predictions in the Case B limit.

The Figure shows relative intensities of the three
brightest Balmer lines. At low densities, the spectra 
depend mainly on temperature, but for densities
$n_{H} > 10^7$ cm$^{-3}$ collisions 
affect the upper levels of the Balmer lines, making the ratios vary through the parameter space. Most galactic nebulae have lower densities, where the line ratios are nearly constant, so we anticipate that their line emission will be near Case B values.

\section {The new physics}
What additional physics might occur and affect the Balmer lines?
We show the effects of an external cool, non-ionizing, radiation
field such as the SED of a population of low-mass stars or red supergiants upon the hydrogen level populations.
We find that continuum pumping by near UV and visible light
can affect the Balmer lines. 
Non-ionizing radiation can photoexcite hydrogen atoms from 
the metastable 2s level, altering the balance of the 
recombination cascade and producing measurable departures
from Case~B predictions. 

\subsection{Model Setup}
D26 showed that most clouds photoionized by a blackbody or stellar
continuum have H~I emission close to Case B. This would apply
to an evolving star cluster, such as the result of a recent epoch of star formation.
Here, we ask whether an additional source of non-ionizing radiation,
such as the SED of an older epoch of star formation, consisting of red super- or hyper-giants, or a number of B stars, can explain the observed departures from Case~B. 

To investigate this possibility in a realistic H~\textsc{ii}-region environment, 
we constructed a set of models using the development version of
\textsc{Cloudy} \citep[see][]{2025RMxAA..61..120G}.
Our baseline model is based on the Orion H~\textsc{ii}-region calculations
originally developed by \citet{1991ApJ...374..580B}. The ionizing continuum is
represented by a TLUSTY O-star atmosphere with an effective temperature of
$40{,}400$~K. We adopt a hydrogen density of
$\log(n_{\rm H}/{\rm cm^{-3}})=2$ and an ionization parameter of
$\log U=-1.5$, together with standard H~\textsc{ii}-region abundances,
Orion-type grains, a microturbulent velocity of $8~{\rm km\,s^{-1}}$, and
cosmic rays at the background level.

We add an additional blackbody (BB) radiation field. This
second continuum is not intended to provide the ionizing photons that create
the H~\textsc{ii} region; rather, it allows us to isolate the effect of a
cooler radiation field on the excited-state populations and Balmer emission of hydrogen.
We vary both the temperature, $T_{\rm eff}$, which determines the spectral
shape of this additional continuum, and the intensity striking the gas.


\subsection{Dependence on the External Radiation-Field's Shape}

\begin{figure*}[ht!]
\plotone{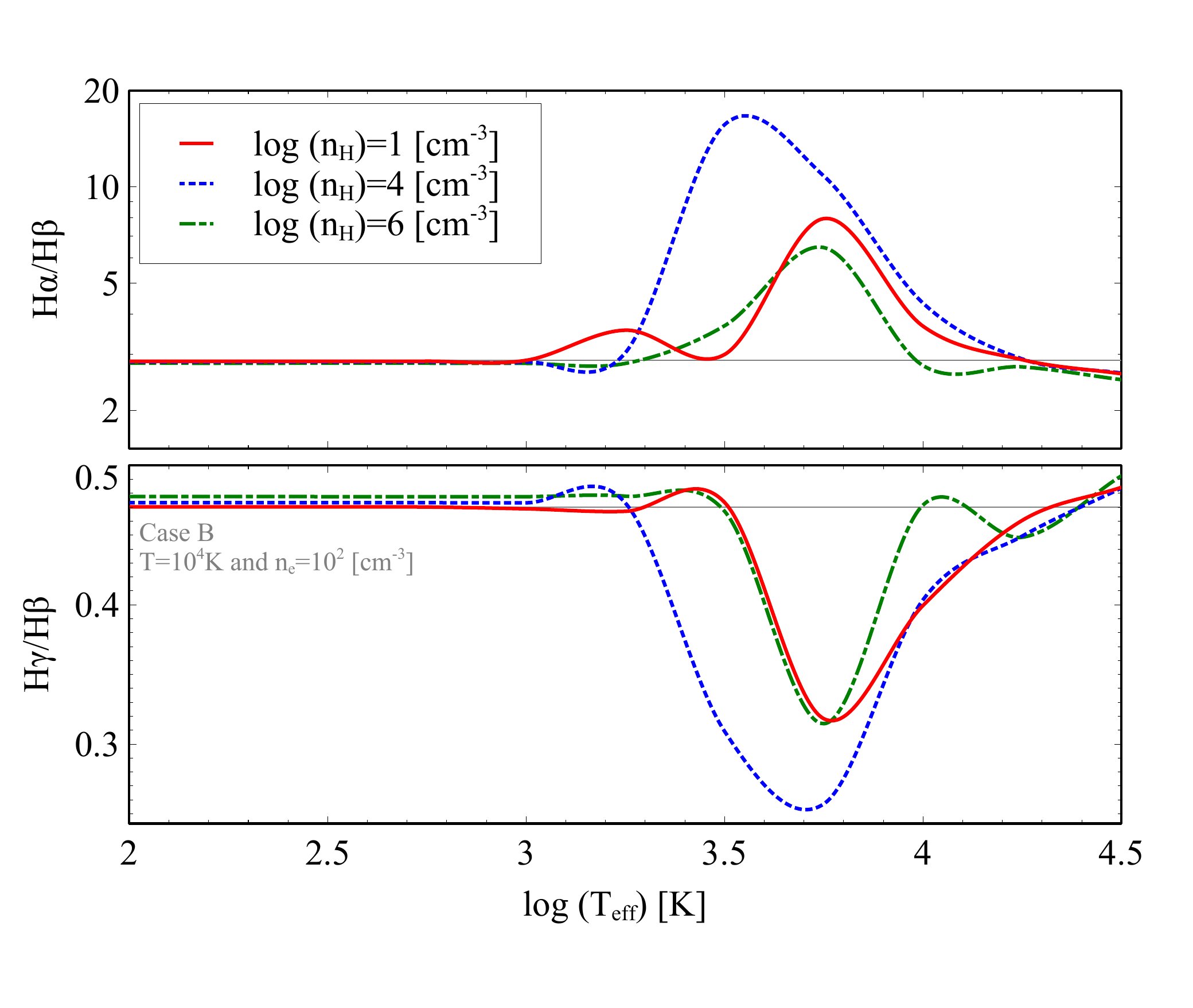}
\caption{H$\alpha$/H$\beta$ (top) and H$\gamma$/H$\beta$ (bottom) as
functions of the temperature of the incident blackbody for three hydrogen
densities. Horizontal lines indicate the Case~B values. 
\label{fig:ratio}}
\end{figure*}

We first ask which photons in the additional radiation field are most
effective at perturbing the Balmer spectrum. We therefore vary
$T_{\rm eff}$ while holding the remaining properties of the nebula fixed.
Figure~\ref{fig:ratio} shows the resulting H$\alpha$/H$\beta$ and
H$\gamma$/H$\beta$ ratios as functions of $T_{\rm eff}$. 

As shown in the Figure, at both low and very high T$_{\rm eff}$, the Balmer ratios remain close to their canonical Case~B values. 
However, at intermediate T$_{\rm eff}$, deviations from Case B are substantial.
The most dramatic feature is a strong enhancement in $H\alpha/H\beta$, accompanied by a suppression of $H\gamma/H\beta$, near
$\log(T_{\rm eff}/{\rm K})\sim3.2-4.2$. As the temperature increases above 10$^{3}$ K, the radiation field becomes efficient at photoexcitation of Balmer lines from the metastable 2s level, 
driving a dramatic increase in H$\alpha$/H$\beta$ while suppressing H$\gamma$/H$\beta$. At still higher temperatures, the balance between the excited 
hydrogen levels changes again, causing H$\alpha$/H$\beta$ to decrease below its Case B 
value while H$\gamma$/H$\beta$ rises above it. This higher-temperature branch is particularly 
important because it moves the Balmer ratios in the same qualitative direction as 
several recent observations. Although the location and amplitude of the strongest 
feature vary somewhat with density, the overall behavior is remarkably similar across 
three orders of magnitude in density. This demonstrates that continuum pumping is 
primarily controlled by the incident radiation field rather than by collisional processes within the nebula.
Figure~\ref{fig:ratio}, therefore, shows that large departures occur over a restricted range of
$T_{\rm eff}$. Next, we separate the effects of the spectral shape and intensity
by independently varying $T_{\rm eff}$ and the dilution factor $W$, defined as the ratio of the local mean intensity to the blackbody intensity at the same temperature.

                                                                                       
\subsection{Dependence on the External Radiation Field's Intensity}

Having identified the range of $T_{\rm eff}$ that produces the largest
changes in the Balmer spectrum, we next ask how the magnitude of these
changes depends on the local intensity of the additional radiation field.
We represent this continuum as a diluted blackbody with an intensity set by the dimensionless dilution factor $W$;
\citep{2014tsa..book.....H,2009MNRAS.398.1809K},
\begin{equation}
    J_\nu = W B_\nu(T_{\rm eff}),
\end{equation}
where $B_\nu(T_{\rm eff})$ is the Planck function. Thus,
\begin{equation}
    W = \frac{J_\nu}{B_\nu(T_{\rm eff})} = \frac{\rm Local\ radiation\ Intensity}{\rm Intensity\ of\ a\ eff\ with\ the\ same\ T_{eff}},
\end{equation}
Thus, $W$ provides a simple measure of the strength of the local radiation
field relative to a blackbody of the same temperature; for example,
$\log W=-2$ corresponds to 1\% of the blackbody intensity while
$W \equiv 1$ corresponds to strict thermodynamic equilibrium.
This is useful when considering the importance of continuum pumping.
It is analogous to the ionization parameter $U$ \citep{2006agna.book.....O}, the dimensionless ratio
of photon to hydrogen densities, a measure of how ionized a gas will become.
For comparison with a physical source, $W$ can be related to its luminosity once a source--cloud distance is specified. 
For a spherical source of radius $R_\star$ a distance $r_0$ away from the gas,
the dilution factor is 
\begin{equation}
W=J_\nu/B_\nu =\frac{L}{16 \pi \sigma T^{4}_{eff}r^{2}_{0}}\simeq \frac{1}{4}
\left(\frac{R_\star}{r_0}\right)^2,\label{eq:dilution}
\end{equation}
 assuming $R_\star \ll r_0$.
This relation provides a direct connection between the parameter $W$ used in
our calculations and the luminosity of a possible source of the additional
radiation field. For a given $T_{\rm eff}$, a value of $W$ specifies the
combination $L/r_0^2$, i.e., the luminosity required at a given distance from
the emitting gas. Thus, $W$ does not correspond to a unique source luminosity
unless the source--cloud separation is specified, just as the ionization parameter $U$ does not specify the stellar luminosity.

In our calculations, $T_{\rm eff}$ and $W$ provide a simple way
to vary two properties of the local radiation field: $T_{\rm eff}$ controls
its spectral shape, while $W$ controls its intensity and hence the rate at which radiative pumping competes with the processes that normally establish the Case~B level populations.   

\begin{figure*}[ht!]
\plotone{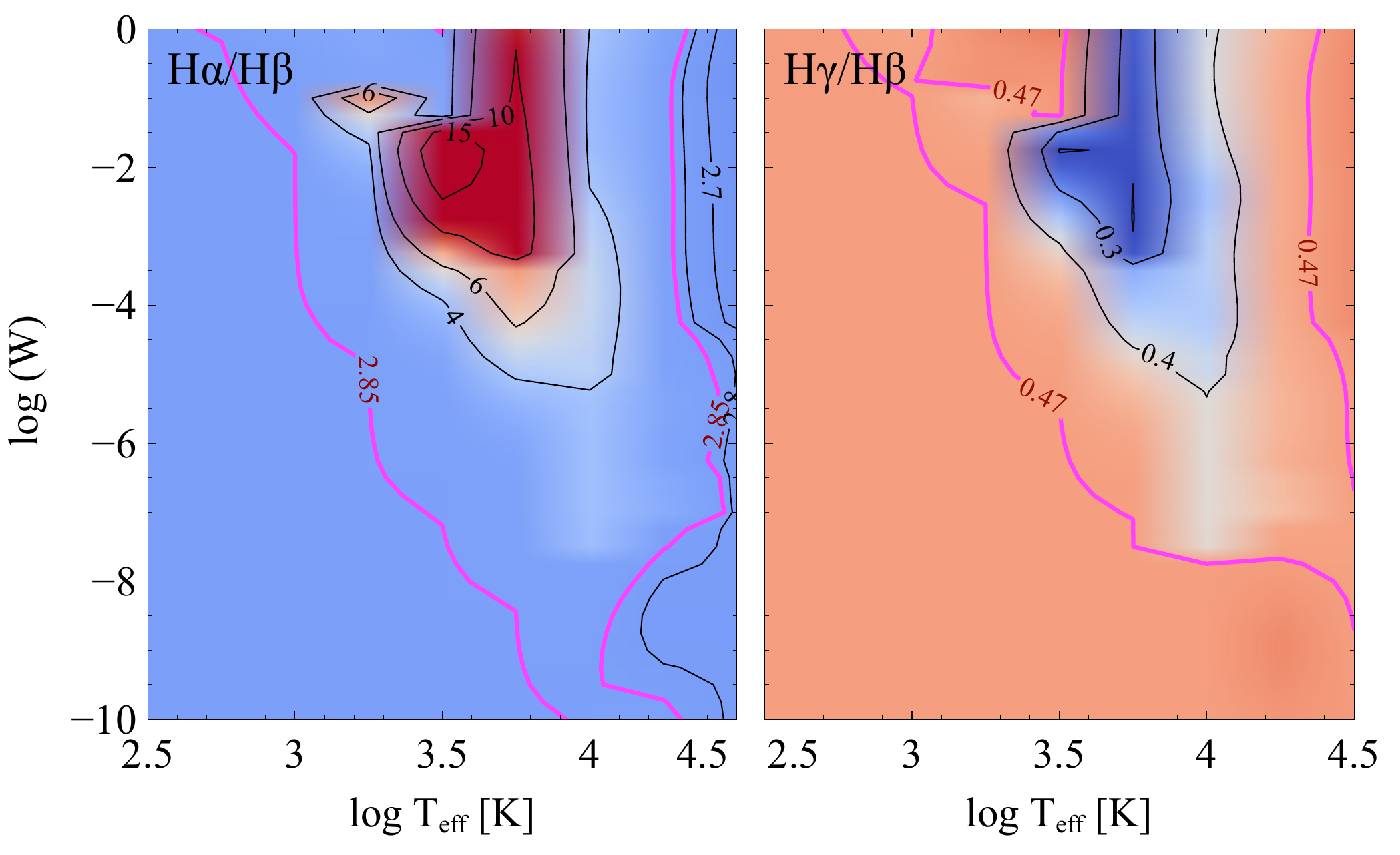} 
\caption{H$\alpha$/H$\beta$ (left) and H$\gamma$/H$\beta$ (right) as
functions of blackbody temperature and dilution factor. Magenta contours
mark the canonical Case~B values.
\label{fig:DCont}}
\end{figure*}

Figure~\ref{fig:DCont} shows the results of changing both the shape and the intensity of the external radiation source. At sufficiently small $W$, the external continuum becomes
dynamically unimportant, and the line ratios approach their Case~B values,
largely independent of $T_{\rm eff}$. As the radiation field becomes stronger,
the temperature-dependent structure identified in Figure~\ref{fig:ratio}
emerges and grows dramatically. 
The largest departures therefore require both the appropriate spectral
shape and sufficient intensity to compete with the normal recombination
cascade. We next investigate the underlying physics.

\begin{figure*}[ht!]
\plotone{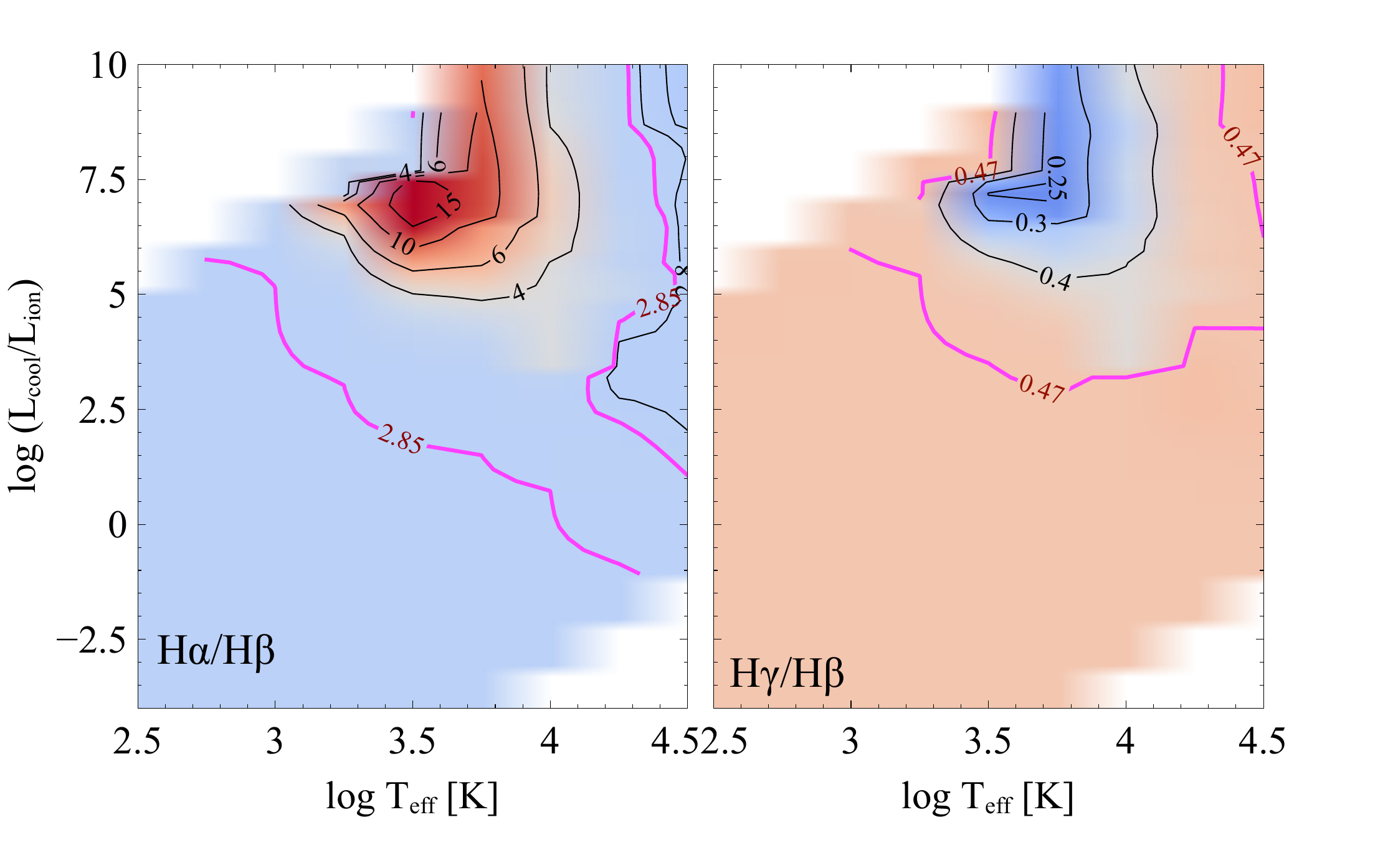} 
\caption{Balmer-line ratios as a function of $T_{\rm eff}$ and the luminosity ratio
$L_{\rm cool}/L_{\rm ion}$. The luminosity ratio is obtained from the
dilution-factor grid assuming co-spatial cool and ionizing sources. Triangles
indicate models extending beyond the sampled grid.
\label{fig:Lratio}}
\end{figure*}

To make the strength of the additional radiation field more physically
transparent, we recast the dilution factor in terms of the luminosity ratio
$L_{\rm cool}/L_{\rm ion}$. In our geometry, the cool and
ionizing sources are necessarily co-spatial and therefore have the same
source--cloud distance, so this distance cancels in the luminosity ratio.
For the diluted blackbody,
\begin{equation}
4\pi J_{\rm cool}=4W\sigma T_{\rm eff}^{4}.
\end{equation}
For the ionizing source, we obtained the corresponding bolometric incident
intensity directly from the Cloudy model by integrating the saved incident
TLUSTY continuum,
\begin{equation}
4\pi J_{\rm ion} =
\int (4\pi\nu J_\nu)_{\rm ion}\,d\ln\nu
=25.91~{\rm erg~cm^{-2}~s^{-1}}.
\end{equation}
The luminosity ratio used below is therefore
\begin{equation}
\frac{L_{\rm cool}}{L_{\rm ion}}
 = \frac{4W\sigma T_{\rm eff}^{4}}{4\pi J_{\rm ion}},
\end{equation}
or, for the adopted ionizing continuum,
\begin{equation}
\log\left(\frac{L_{\rm cool}}{L_{\rm ion}}\right)
 = \log W +4\log(T_{\rm eff}/{\rm K})-5.058. \label{eq:Lratio}
\end{equation}
We verified this conversion independently by integrating a Cloudy model
containing only the diluted blackbody; the resulting incident intensity agrees
with $4W\sigma T_{\rm eff}^{4}$.
This conversion should be understood within the limitations of our idealized
geometry. Placing the cool and ionizing sources at different
positions within the nebula would require a multidimensional radiative-transfer
calculation. The luminosity ratios quoted here therefore correspond specifically
to the co-spatial source geometry modeled by Cloudy. Figure~\ref{fig:Lratio} shows the results. The vertical axis gives the luminosity ratio of the additional cool component
to the ionizing TLUSTY source, calculated from the dilution-factor grid using
Eq.~(\ref{eq:Lratio}). Because this figure is obtained by transforming the
finite $T_{\rm eff}$--$W$ grid shown in Figure~3, parts of the transformed
parameter space are not sampled.  Upward and downward triangles indicate values for which the corresponding model lies beyond the upper or lower boundary of the original dilution-factor grid.

Figure~\ref{fig:Lratio} also illustrates an important limitation of this geometry.
For the co-spatial configuration assumed here, significant departures
from Case~B require very large values of $L_{\rm cool}/L_{\rm ion}$.
However, more generally,
\begin{equation}
\left(\frac{L_{\rm cool}}{L_{\rm ion}}\right)
=
\left(\frac{L_{\rm cool}}{L_{\rm ion}}\right)_{\rm co-spatial}
\left(\frac{r_{\rm cool}}{r_{\rm ion}}\right)^2 .
\end{equation}
Thus, a cool source located substantially closer to the Balmer-emitting gas
than the ionizing source could produce the same pumping with a much smaller
intrinsic luminosity. Such a configuration would also introduce additional physics absent from our idealized static models, including stellar winds and shocks. A self-consistent treatment of these effects, together with the multidimensional radiation field is beyond the scope of this Letter.
\subsection{What Is Going On?}
The underlying mechanism is radiative excitation of hydrogen by the external continuum 
field. The cool blackbody continua responsible for the strongest pumping signature 
contain relatively little hydrogen-ionizing radiation, but can nevertheless strongly 
modify the excited-state populations through continuum pumping and resonant radiative 
coupling. The balance among recombination, collisional excitation, radiative decay, 
ionization, and radiative pumping determines the hydrogen-level populations
\citep{2006agna.book.....O}. However, 
the external continuum can alter this balance by populating excited levels 
and modifying the relative strengths of the cascade pathways. Such cool blackbodies cannot excite Lyman lines with their large energies but can excite Balmer lines from the metastable 2s level. Since the Balmer lines 
originate from different upper levels, even relatively modest changes in the excitation balance can produce measurable changes in the observed line ratios.

The effect is strongly temperature-dependent because changing \(T_{\rm eff}\) changes the number of photons available at the energies relevant for photoexcitation from the \(n=2\) level. In the calculations presented below, the dominant process is photoexcitation from the metastable 2s level. Excitation to levels with \(n>3\) also opens additional cascade pathways through Paschen and higher-series transitions.


The blackbodies considered here should therefore not be interpreted literally
as co-spatial stellar populations with the luminosities implied by Figure~\ref{fig:Lratio}.
Rather, they parameterize the local soft radiation field required to produce
the pumping. A physically plausible realization may instead require a geometry
in which a cool source lies much closer to the Balmer-emitting gas than the
dominant ionizing source. Most sources have H~{\sc i} emission that can be
explained by Case~B \citep [for example see][]{2025MNRAS.540..190M}, while significant
departures appear to occur only in a small fraction of systems. This suggests
that the conditions required to produce non-Case~B emission are themselves
unusual. One possibility explored here is that these systems contain a strong
local cool radiation field, perhaps associated with a distinct stellar
population and star-formation episode. If so, the anomalous Balmer spectrum
could provide information not only about the local radiation field and
geometry, but also about the star-formation history of these distant galaxies.

\subsection{Implications for High-Redshift Systems (JWST observations)}


At high redshift, many systems likely contain complex mixtures of stellar
populations, including evolved or low-mass stars that contribute substantial
optical and infrared continuum radiation while producing relatively little
ionizing flux. Such populations may be difficult to identify directly,
particularly in unresolved systems where their continuum is blended with
that of younger, more luminous stars. Under favorable conditions, their
imprint on the Balmer decrement may therefore provide one of the few, or
even the only, observable signatures of their presence.
If such radiation
fields coexist with compact H~\textsc{ii} regions, continuum 
pumping may modify the local hydrogen level populations and alter 
the emergent Balmer spectrum.

\begin{figure}[ht!]
\plotone{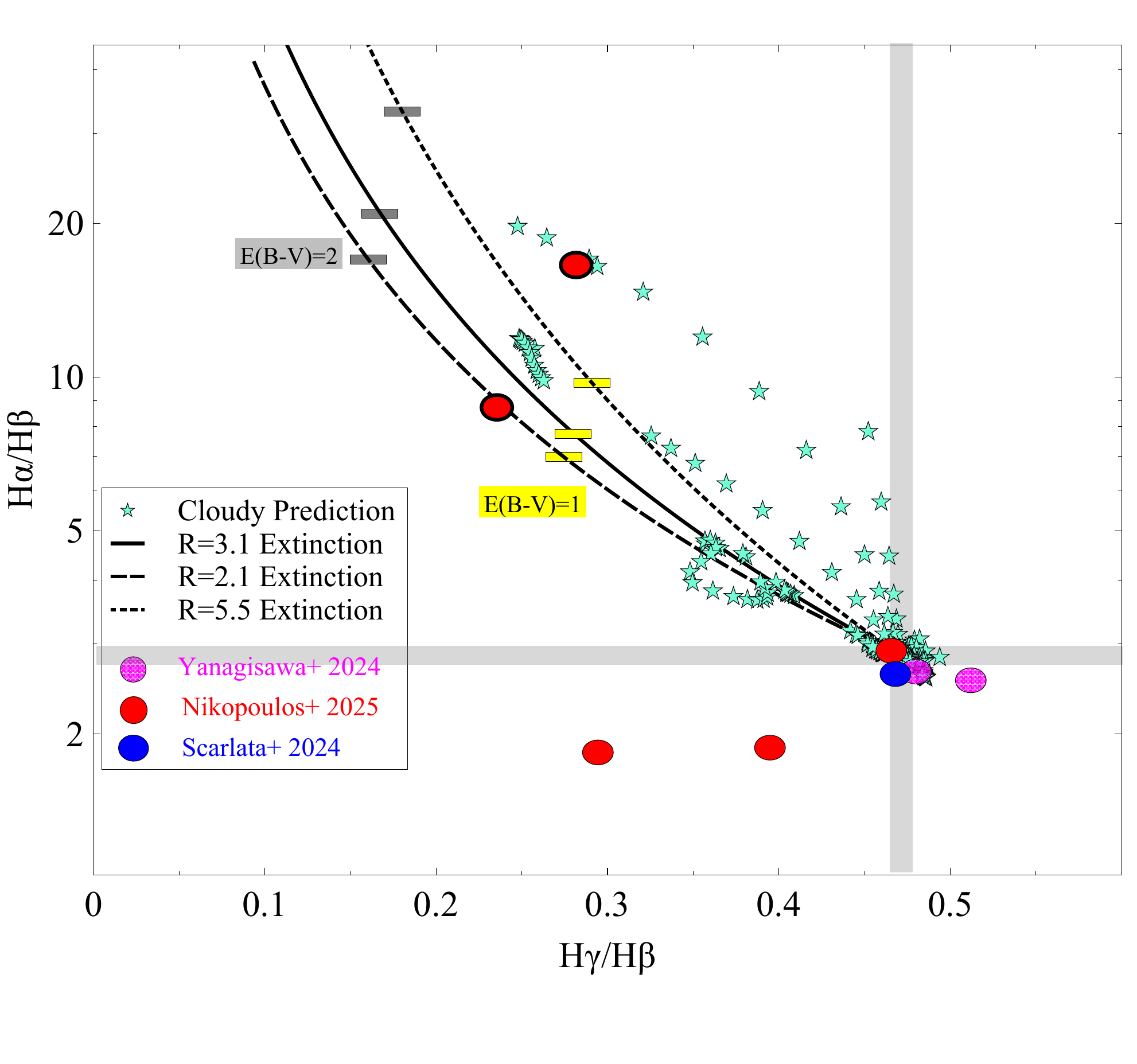}
\caption{H$\alpha$/H$\beta$ versus H$\gamma$/H$\beta$ for our Cloudy
models (shown and described in Fig.\ref{fig:DCont} and its text) and published measurements. Curves show dust-reddening tracks for
$R_V=2.1$, 3.1, and 5.5; gray bands indicate the Case~B range for
$5000<T_{\rm e}<10^4$~K at $n_{\rm H}=10^2~{\rm cm^{-3}}$. Note that For \cite{2025arXiv251006362N}, thick-outlined circles indicate the broad'' components, while normal-outlined circles indicate the narrow'' components.} 
\label{fig:DScat}
\end{figure}

Figure~\ref{fig:DScat} compares our Cloudy predictions (resulting from the same models as in Fig~\ref{fig:DCont}) with several published measurements. 
The extinction tracks follow the \cite{1989ApJ...345..245C} extinction law
and were calculated with \texttt{python} \citep{python}.
They occupy a relatively narrow locus in the H$\alpha$/H$\beta$--
H$\gamma$/H$\beta$ plane, indicating that the precise choice of extinction curve has only a modest effect on the expected 
trajectory of dust reddening. By contrast, many of our models 
populate regions that lie well away from any reddening sequence. In
particular, continuum pumping by cool, non-ionizing radiation 
fields produces combinations of H$\alpha$/H$\beta$ and 
H$\gamma$/H$\beta$ that cannot be reached by dust extinction alone. 
So, while reddening moves Balmer ratios along a nearly one-
dimensional path, continuum pumping shifts them into a distinct region of parameter space.

This distinction has important implications for interpreting Balmer decrements in 
distant galaxies. Anomalous hydrogen-line ratios are often attributed exclusively to dust extinction or, in extreme 
cases, to very high gas densities or Balmer line optical depths. Our calculations self-consistently include line optical depths and therefore allow for Balmer-line opacity,
but demonstrate that an additional mechanism is also possible: non-ionizing continuum 
radiation can modify the hydrogen excitation balance and produce characteristic departures from the classical Case B locus.
Therefore, simultaneous measurements of H$\alpha$ / H$\beta$ and H$\gamma$ / H $\beta$ provide a powerful 
diagnostic to distinguish between reddening and intrinsic 
changes in the hydrogen recombination cascade. 
These deviations may serve as indirect tracers of cool, unresolved stellar populations that contribute little ionizing radiation but leave a measurable imprint on the emergent Balmer spectrum. In this picture, anomalous Balmer decrements are not simply failures of Case~B recombination; they are the spectroscopic footprints of the radiation field experienced by the hydrogen atom.

Figure~~\ref{fig:DScat} also makes clear that continuum pumping by a cool radiation field cannot explain the full diversity of the observed Balmer ratios. In particular, some of the observed systems with unusually low H$\alpha$/H$\beta$ lie well outside the region covered by our models, suggesting that additional physics is involved. The cool radiation field explored here therefore represents only one of several possible mechanisms for producing departures from Case~B. In future work, we will investigate other physical scenarios that may help bridge the remaining gap between theoretical predictions and the growing body of JWST observations.
\section{Conclusions}
Most galactic emission nebulae have H~I spectra close to Case~B, yet some of the largest departures reported to date have been found in a few high-redshift systems \cite[e.g. ][]{2024ApJ...974..180Y,2025MNRAS.540..190M, 2025arXiv251006362N}. What physics is different about these special environments?
This paper explores continuum pumping by a cool, non-ionizing radiation field as an
additional mechanism for producing anomalous Balmer decrements. Our
calculations 
show that, although Case~B is remarkably robust under ordinary nebular
conditions, a strong cool continuum can significantly modify the Balmer
spectrum, but would have few other effects. 

Together, these calculations reveal a clear progression. Case~B remains a good description of H~I emission under
conditions representative of nearby nebulae (Figure~1). Here, we ask whether
an additional radiation field can provide the new physics needed to explain
the observed departures from Case~B. We find that its effect depends strongly
on which photons are available, as demonstrated by varying $T_{\rm eff}$
(Figure~2), and on the intensity of the radiation field, as shown by varying
both $T_{\rm eff}$ and $W$ (Figure~\ref{fig:DCont}). The resulting Balmer ratios can occupy
regions of the H$\alpha$/H$\beta$--H$\gamma$/H$\beta$ plane that cannot be
reached by dust reddening alone (Figure~\ref{fig:DScat}), providing a potential
observational signature of otherwise hidden radiation fields.

The largest departures occur at around
$\log(T_{\rm eff}/{\rm K})\sim3.2$--$4.2$. Photoexcitation from excited
H~\textsc{i}, particularly the metastable $2s$ level, modifies the hydrogen
level populations and subsequent cascade. Both the spectral shape and
intensity of the incident continuum determine the magnitude of the effect.
This mechanism can produce steep H$\alpha$/H$\beta$ ratios, but does not
account for the unusually low H$\alpha$/H$\beta$ ratios observed in some
systems, where additional physics must be involved.
Importantly, continuum pumping and dust reddening follow different
trajectories in the H$\alpha$/H$\beta$--H$\gamma$/H$\beta$ plane.
Multiple Balmer lines can therefore distinguish between these effects
and potentially reveal a strong local soft radiation field that is
otherwise difficult to detect. Our atomic models show
the atomic-physics effect, but the large luminosity ratios implied by
the co-spatial geometry suggest that physically realistic applications
may require the cool source to lie substantially closer to the
Balmer-emitting gas than the dominant ionizing source. In unresolved
systems, anomalous Balmer decrements may therefore provide a signature
of otherwise hidden components of the local radiation field.

All input files used to generate the models and figures presented in this work 
are publicly available in the Cloudy project repository.\footnote{
\url{https://gitlab.nublado.org/cloudy/papers/CaseB-External-source-D26}}

\begin{acknowledgments}
We thank NASA for support through grants 24-APRA24-012, 24-XGO24-0001, as well as  JWST-AR-06419, JWST-AR-06428,
JWST-GO-5018, JWST-GO-07033, and JWST-GO-5354 from the Space Telescope Science Institute. We also acknowledge support from the NSF
under Grant No. 2606041.
\end{acknowledgments}

\software{  
          Cloudy \citep{2025RMxAA..61..120G}, Veusz \citep{2023ascl.soft07017S}, Python \citep{software-citation-station-paper,software-citation-station-zenodo}
          }



\bibliography{caseB}{}
\bibliographystyle{aasjournalv7}



\end{document}